# Optimal teleportation via thermal entangled states of a two-qubit Heisenberg Chain

Yue Zhou[1], Guo-Feng Zhang[2][*] and Shu-Shen Li[1], Ahmad Abliz[3]

[1]*State Key Laboratory for Superlattices and Microstructures, Institute of Semiconductors, Chinese Academy of Sciences, P. O. Box 912, Beijing 100083, China*

[2] *School of Physics, Beijing University of Aeronautics & Astronautics, Xueyuan Road No. 37, Beijing 100083, China*

[3] *School of Mathematics, Physics and Informatics, Xinjiang Normal University, Urumchi 830054, China*

**Abstract:** We study the optimal teleportation based on Bell measurements via the thermal states of a two-qubit Heisenberg *XXX* chain in the presence of Dzyaloshinsky-Moriya (DM) anisotropic antisymmetric interaction and obtain the optimal unitary transformation. The explicit expressions of the output state and the teleportation fidelity are presented and compared with those of the standard protocol. It is shown that in this protocol the teleportation fidelity is always larger and unit fidelity is achieved at zero temperature. The DM interaction can enhance the teleportation fidelity at finite temperatures, as opposed to the effect of the interaction in the standard protocol. Cases with other types of anisotropies are also discussed.

PACS numbers: 03.67.Hk, 75.10.Jm

Quantum teleportation is a fascinating phenomenon based on the nonlocal properties of quantum mechanics and plays an important role in quantum information. During teleportation, the unknown teleported state in the sender's qubit is destroyed and then reconstructed in the target qubit, the whole process relies on local unitary transformations and some assistance of classical communication. In order to realize quantum teleportation, an entangled quantum channel is a prerequisite. The original protocol proposed by Bennett *et al*. [1] relies on two particles prepared in an Einstein-Podolsky-Rosen (EPR) singlet state. Bowen and Bose [2] showed that standard teleportation with an arbitrary entangled mixed state resource is equivalent to a generalized depolarizing channel with probabilities given by the maximally entangled components of the resource.

Of the several sources proposed, the one based on Heisenberg interaction in solid state systems is an attractive field. The state of a typical solid state system **in thermal equilibrium at temperature** $T$ is described by $\rho(T) = e^{-H/kT}/Z$, where $H$ is the Hamiltonian, $Z = tr e^{-H/kT}$ is the partition function and $k$ is the Boltzmann's constant. The entanglement associated with a thermal equilibrium state $\rho(T)$ is called thermal entanglement. Compared with other kinds of entanglement, the thermal entanglement is more immune to various sources of decoherence and requires neither measurement nor controlled switching of interactions in the preparation process. In view of the above advantages, thermal entanglement has gained much interest recently [3-8], and teleportation via **thermally entangled states** of two-qubit Heisenberg *XX* [9], *XY* [10], *XXX* [11], *XXZ* [12], and *XYZ* [13] chains in the presence of an inhomogeneous magnetic field [12,13] and a Dzyaloshinsky-Moriya (DM) interaction [11, 13] has been intensively studied by several authors. However, the investigations were based exclusively on the standard protocol; that is to say, during the teleportation process **first a joint Bell measurement and then Pauli rotations on the target qubit depending on the measurement result are performed**. In this paper, we consider the optimal teleportation based on the Bell measurement [14]. Instead of Pauli rotations, in this protocol one tries his best to choose a particular unitary transformation depending on the quantum channel, in order to obtain the maximal teleportation fidelity. The thermal entanglement depends wholly on the temperature and the intrinsic properties of the system, which are controllable or measurable beforehand, so it is well suitable for this protocol. The paper is organized as follows: **first**, we investigate the process of optimal teleportation via a two-qubit Heisenberg *XXX* chain with DM interaction; **then**, we compare the result with that of the standard protocol; **and lastly**, we briefly discuss cases with other kinds of anisotropies.

---

[*] Corresponding author; electronic mail address: gf1978zhang@buaa.edu.cn

The Hamiltonian of a two-qubit Heisenberg *XXX* chain in presence of DM interaction is given by

$$H = \frac{1}{2}[J\boldsymbol{\sigma}_1 \cdot \boldsymbol{\sigma}_2 + \boldsymbol{D} \cdot (\boldsymbol{\sigma}_1 \times \boldsymbol{\sigma}_2)], \tag{1}$$

where $J$ is the real exchange coupling coefficient, the model is called antiferromagnetic for $J > 0$ and ferromagnetic for $J < 0$. Because the teleportation is generally not realizable for the ferromagnetic cases [11], only the antiferromagnetic cases are considered here. $\boldsymbol{D} \cdot (\boldsymbol{\sigma}_1 \times \boldsymbol{\sigma}_2)$ is the DM coupling term that arises from spin-orbit coupling [15, 16]. **Because all the items except $\boldsymbol{D}$ in eq. (1) are rotation invariant, without loss of generality, we can choose $\boldsymbol{D} = D\boldsymbol{z}$, then** the Hamiltonian (1) becomes

$$H = \frac{1}{2}[J(\sigma_1^x\sigma_2^x + \sigma_1^y\sigma_2^y + \sigma_1^z\sigma_2^z) + D(\sigma_1^x\sigma_2^y - \sigma_1^y\sigma_2^x)] \tag{2}$$

The eigenstates and corresponding eigenvalues of the Hamiltonian (2) can be expressed as

$$|\phi_1\rangle = |00\rangle, \qquad E_1 = \frac{J}{2},$$

$$|\phi_2\rangle = |11\rangle, \qquad E_2 = \frac{J}{2},$$

$$|\phi_3\rangle = \frac{\sqrt{2}}{2}(|10\rangle + e^{i\vartheta}|01\rangle), \qquad E_3 = -\frac{J}{2} + \sqrt{J^2 + D^2},$$

$$|\phi_4\rangle = \frac{\sqrt{2}}{2}(|10\rangle - e^{i\vartheta}|01\rangle), \qquad E_4 = -\frac{J}{2} - \sqrt{J^2 + D^2}, \tag{3}$$

where $\vartheta = -\arctan(D/J)$, while $|1\rangle$ and $|0\rangle$ denote the spin-up and spin-down states, respectively.

In thermal equilibrium state at temperature $T$, the density matrix $\rho(T)$ of the system is given by

$$\rho(T) = \frac{1}{Z}e^{-H/kT} = \frac{1}{Z}\begin{pmatrix} e^{-J/2T} & 0 & 0 & 0 \\ 0 & e^{J/2T}\cosh\omega & -e^{-i\vartheta}e^{J/2T}\sinh\omega & 0 \\ 0 & -e^{i\vartheta}e^{J/2T}\sinh\omega & e^{J/2T}\cosh\omega & 0 \\ 0 & 0 & 0 & e^{-J/2T} \end{pmatrix} \tag{4}$$

in the standard basis $\{|11\rangle, |10\rangle, |01\rangle, |00\rangle\}$, where $Z = 2e^{-J/2T} + 2e^{J/2T}\cosh(\sqrt{J^2 + D^2}/T)$ and $\omega = \sqrt{J^2 + D^2}/T$. Note that we have set $k = 1$ in eq. (4) and in all following equations for simplification.

According to the physical meaning of the thermal equilibrium state, it is easy to decompose $\rho(T)$ into

$$\rho(T) = \sum_{i=1}^{4} p_i |\phi_i\rangle\langle\phi_i|, \tag{5}$$

where $p_i$ **are the probability distribution and satisfy** $p_i = Z^{-1}e^{-E_i/T}$. The entanglement property of $\rho(T)$ primarily depends on the ground state $|\phi_4\rangle$, which is a maximal entangled state. At zero temperature, $p_4 = 1$ and only the ground state $|\phi_4\rangle$ is populated; while at low temperatures $|\phi_4\rangle$ takes a dominant part, so that $\rho(T)$ is entangled. With the increase of the temperature, the three excited states are mixed into the ground state, so the entanglement degree of $\rho(T)$ decreases and vanishes above the critical temperature. Because $p_4 = Z^{-1}e^{-E_4/T}$, the ground state $|\phi_4\rangle$ is more populated if $E_4$ is low, which leads to a larger degree of entanglement at the same finite temperature.

Quantum teleportation via an arbitrary mixed state was first investigated by Bowen and Bose [2]. **They showed that, when an arbitrary two-qubit mixed state** $\chi$ **was used as the resource, the teleported state in the standard teleportation protocol** $T_0$ **is given by**

$$\Lambda_{T_0}\rho = \sum_i Tr[E_i \chi]\sigma^i \rho \sigma^i, \tag{6}$$

where $E_i = \sigma^i|\psi^+\rangle\langle\psi^+|\sigma^i$ and $\sigma^0 = I$, $\sigma^1 = \sigma^x$, $\sigma^2 = \sigma^y$ and $\sigma^3 = \sigma^z$. And later, Gu et al. further generalized the above result [17]. During their derivation, they adopted a superposition of all the output states corresponding to each outcome of the joint Bell-basis measurement. In the practical teleportation process, however, one achieves the teleported state corresponding to only one of the possible measurement outcome, instead of a superposition of them. In addition, the teleported states corresponding to different measurement outcomes may not be identical. In order to investigate the optimal unitary transformation, it is necessary to find the explicit expression of the output state corresponding to each measurement outcome. So here we derive them from a different approach. Suppose the state to be teleported is given by $|\varphi_A\rangle = \cos\frac{\theta}{2}|1\rangle + e^{i\delta}\sin\frac{\theta}{2}|0\rangle$ ($0 \leq \theta \leq \pi$, $0 \leq \delta \leq 2\pi$), then the joint **three-spin** system is in the product state given by

$$\rho_c = \rho_A \otimes \rho(T), \tag{7}$$

where $\rho_A = |\varphi_A\rangle\langle\varphi_A|$ **is the density matrix corresponding to the input state. First we define the projection operators** $M_i$:

$$M_i = |\psi_i\rangle\langle\psi_i| \otimes I. \tag{8}$$

where $|\psi_1\rangle = |\psi^-\rangle = 2^{-1/2}(|10\rangle - |01\rangle)$, $|\psi_2\rangle = |\psi^+\rangle = 2^{-1/2}(|10\rangle + |01\rangle)$, $|\psi_2\rangle = |\phi^-\rangle = 2^{-1/2}(|11\rangle - |00\rangle)$

and $|\psi_4\rangle = |\phi^+\rangle = 2^{-1/2}(|11\rangle + |00\rangle)$ are the four Bell states. **Due to the completeness of $\{|\psi_i\rangle\}$, there is**

$$\sum_i M_i = \sum_i |\psi_i\rangle\langle\psi_i| \otimes I = I^{\otimes 3}. \tag{9}$$

**So we have**

$$\begin{aligned}\rho_c &= \sum_i M_i \rho_c \sum_j M_j^\dagger \\ &= \sum_i p_i |\psi_i\rangle\langle\psi_i| \otimes \rho_i + \sum_{ij}{}' p_{ij} |\psi_i\rangle\langle\psi_j| \otimes \rho_{ij} \end{aligned} \tag{10}$$

When a joint Bell-basis measurement is performed on the first two spins, the state of the third spin (the second spin in eq. (2)) will collapse into one of the four states $\rho_1$, $\rho_2$, $\rho_3$ or $\rho_4$. **Performing $M_i$ onto $\rho_c$ yields**

$$M_i \rho_c M_i^\dagger = p_i |\psi_i\rangle\langle\psi_i| \otimes \rho_i. \tag{11}$$

**Then we can obtain the unnormalized $\rho_i$ by tracing over the first two qubits**

$$\rho_i' = p_i \rho_i = tr_{12}(M_i \rho_c M_i^\dagger). \tag{12}$$

**Note that $tr\rho_i' = p_i tr\rho_i = p_i$, hence**

$$\rho_i = \frac{\rho_i'}{tr\rho_i'}. \tag{13}$$

**Make use of eq. (13), the expressions of $\rho_i$ can be obtained straightforwardly**

$$\rho_1 = \frac{e^{J/2T}}{Z} \begin{pmatrix} 2\cosh\omega\cos^2\frac{\theta}{2} + e^{-J/T} & \sinh\omega\sin\theta e^{-i\delta} e^{i\vartheta} \\ \sinh\omega\sin\theta e^{i\delta} e^{-i\vartheta} & 2\cosh\omega\sin^2\frac{\theta}{2} + e^{-J/T} \end{pmatrix},$$

$$\rho_2 = \frac{e^{J/2T}}{Z} \begin{pmatrix} 2\cosh\omega\cos^2\frac{\theta}{2} + e^{-J/T} & -\sinh\omega\sin\theta e^{-i\delta} e^{i\vartheta} \\ -\sinh\omega\sin\theta e^{i\delta} e^{-i\vartheta} & 2\cosh\omega\sin^2\frac{\theta}{2} + e^{-J/T} \end{pmatrix},$$

$$\rho_3 = \frac{e^{J/2T}}{Z} \begin{pmatrix} 2\cosh\omega\sin^2\frac{\theta}{2} + e^{-J/T} & \sinh\omega\sin\theta e^{i\delta} e^{i\vartheta} \\ \sinh\omega\sin\theta e^{-i\delta} e^{-i\vartheta} & 2\cosh\omega\cos^2\frac{\theta}{2} + e^{-J/T} \end{pmatrix},$$

$$\rho_4 = \frac{e^{J/2T}}{Z} \begin{pmatrix} 2\cosh\omega\sin^2\frac{\theta}{2} + e^{-J/T} & -\sinh\omega\sin\theta e^{i\delta} e^{i\vartheta} \\ -\sinh\omega\sin\theta e^{-i\delta} e^{-i\vartheta} & 2\cosh\omega\cos^2\frac{\theta}{2} + e^{-J/T} \end{pmatrix}. \tag{14}$$

Note that the density matrix of the input state is given by

$$\rho_A = \begin{pmatrix} \cos^2\frac{\theta}{2} & \frac{1}{2}\sin\theta e^{-i\delta} \\ \frac{1}{2}\sin\theta e^{i\delta} & \sin^2\frac{\theta}{2} \end{pmatrix}. \tag{15}$$

Comparing eq. (14) with eq. (15), we can see that $\rho_1$ tends to $\rho_A$ while $T$ tends to zero. So at low temperatures, $\rho_1$ per se is the desired state except for some distortion. The distortion arises from the difference of the ground state $|\phi_4\rangle$ to the principal singlet state $|\psi^-\rangle$, as well as from the mixture of the excited states. In the standard protocol, the Pauli rotations $\sigma^z$, $\sigma^x$, and $\sigma^y$ should be applied on $\rho_2$, $\rho_3$, and $\rho_4$, respectively, which gives the output state:

$$\rho_{out} = \frac{e^{J/2T}}{Z}\begin{pmatrix} 2\cosh\omega\cos^2\frac{\theta}{2}+e^{-J/T} & \sinh\omega\sin\theta e^{-i\delta}e^{\pm i\vartheta} \\ \sinh\omega\sin\theta e^{i\delta}e^{\mp i\vartheta} & 2\cosh\omega\sin^2\frac{\theta}{2}+e^{-J/T} \end{pmatrix}. \tag{16}$$

It can be seen from eq. (16) that if $\vartheta$, which arises from the DM interaction, is nonvanishing, the output states corresponding to different Bell measurement outcomes have a small difference in the phase. The upper **signs** of $i\vartheta$ in eq. (16) **correspond** to the outcomes $|\psi^\pm\rangle$ while the **lower ones correspond** to the outcomes $|\phi^\pm\rangle$.

The quality of a teleportation is generally characterized by the fidelity [18], which is essentially a measurement of the distance between the input and the output states. If either of the two states is pure, the fidelity can be written as

$$F = \langle\psi|\rho|\psi\rangle, \tag{17}$$

where $|\psi\rangle$ is the wave vector of the pure state and $\rho$ is the density matrix of the mixed one. The two states are orthogonal for $F=0$ and equivalent for $F=1$. Making use of eq. (17), the fidelity of the teleportation in the standard protocol can be expressed as

$$F_s = \langle\varphi_A|\rho_{out}|\varphi_A\rangle = \frac{e^{J/2T}}{Z}(\cosh\omega + e^{-J/T} + \cosh\omega\cos^2\theta + \sinh\omega\sin^2\theta\cos\vartheta). \tag{18}$$

Then we consider whether there exists a particular unitary transformation that maximize the teleportation fidelity. Besides a negligible global phase, a universal unitary transformation can be written in the form of combination of three rotation operations:

$$U = R_z(\alpha)R_y(\beta)R_z(\gamma) = \begin{pmatrix} e^{-i(\alpha+\gamma)/2}\cos\frac{\beta}{2} & e^{-i(\alpha-\gamma)/2}\sin\frac{\beta}{2} \\ e^{i(\alpha-\gamma)/2}\sin\frac{\beta}{2} & e^{i(\alpha+\gamma)/2}\cos\frac{\beta}{2} \end{pmatrix}, \tag{19}$$

where $\alpha$, $\beta$ and $\gamma$ are the rotation angles. Performing the unitary transformation on $\rho_1$ yields

$$\rho_{out} = U\rho_1 U^\dagger ; \tag{20}$$

then the teleportation fidelity takes the form

$$F = \frac{e^{J/2T}}{Z}[\cosh\omega + e^{-J/T} + \cosh\omega\cos\beta\cos^2\theta + \cos(\vartheta - \alpha - \gamma)\cos^2\frac{\beta}{2}\sinh\omega\sin^2\theta + \frac{1}{2}\cos(\alpha - \delta)\sin\beta\cosh\omega\sin(2\theta) - \frac{1}{2}\cos(\gamma + \delta - \vartheta)\sin\beta\sinh\omega\sin(2\theta) - \cos(\gamma - \alpha - \vartheta + 2\delta)\sin^2\frac{\beta}{2}\sinh\omega\sin^2\theta]. \tag{21}$$

The fidelity of teleportation dependents on the input states, **whose** parameters $\theta$ and $\delta$ are unknown. So we first take an average of the fidelity over the surface of the Bloch sphere:

$$F_a = \frac{\int_0^{2\pi} d\delta \int_0^\pi F\sin\theta d\theta}{4\pi}, \tag{22}$$

which yields

$$F_a = \frac{e^{J/2T}}{Z}[\cosh\omega + e^{-J/T} + \frac{1}{3}\cosh\omega\cos\beta + \frac{2}{3}\cos(\vartheta - \alpha - \gamma)\cos^2\frac{\beta}{2}\sinh\omega] . \tag{23}$$

Because $\cosh\omega > \sinh\omega > 0$, the relations $\cos\beta = 1$ and $\cos(\vartheta - \alpha - \gamma) = 1$ have to hold true in order to maximize the average fidelity. Then, for simplicity, the rotation angles can be chosen as $\alpha = \vartheta$ and $\beta = \gamma = 0$. It follows that the optimal unitary transformation for $\rho_1$ is given by

$$U_o(\vartheta) = \begin{pmatrix} e^{-i\vartheta/2} & 0 \\ 0 & e^{i\vartheta/2} \end{pmatrix}, \tag{24}$$

which is a rotation operation around the z axis and the angle is determined by the ratio of $D$ and $J$. Comparing the expressions of $\rho_2$, $\rho_3$ and $\rho_4$ with that of $\rho_1$, it is seen that the optimal unitary transformations for them are $U_o(\vartheta)\sigma^z$, $U_o(-\vartheta)\sigma^x$ and $U_o(-\vartheta)\sigma^y$, respectively. And the output state is given by

$$\rho_{out} = \frac{e^{J/2T}}{Z}\begin{pmatrix} 2\cosh\omega\cos^2\frac{\theta}{2} + e^{-J/T} & \sinh\omega\sin\theta e^{-i\delta} \\ \sinh\omega\sin\theta e^{i\delta} & 2\cosh\omega\sin^2\frac{\theta}{2} + e^{-J/T} \end{pmatrix}, \tag{25}$$

which is uniform for all the Bell-basis measurement outcomes. Making use of eq. (17), the optimal teleportation fidelity can be obtained as

$$F_o = \frac{e^{J/2T}}{Z}(\cosh\omega + e^{-J/T} + \cosh\omega\cos^2\theta + \sinh\omega\sin^2\theta) = \frac{e^{J/2T}}{Z}(e^{-J/T} + e^{\omega} + e^{-\omega}\cos^2\theta). \qquad (26)$$

Comparing eq. (26) with eq. (18), we can see that when $\cos\vartheta = 1$, $F_o = F_s$. Thus, the inequality $F_o \geq F_s$ holds true for any values of $\vartheta$. That is, the modified protocol has a larger fidelity not only in average, but also for each individual input state. In the above discussion we assume that the input state is pure, however, the optimal unitary transformation depends only on the quantum channel, so the result is also valid for any mixed input states due to the linearity of the teleportation protocol.

Fig. (1) demonstrates the dependence of the average fidelity of both protocols on the temperature $T$, where $F_{sa}$ and $F_{oa}$ represent average fidelity of the standard and optimal protocol, respectively. Both decrease monotonously as $T$ increases, the curves of $F_{sa}$ and $F_{oa}$ are similar in shape, but the value of $F_{oa}$ is always larger than that of $F_{sa}$. This is especially true when the temperature is low. When the temperature tends to zero, the value of $F_{oa}$ tends to one and the quantum teleportation is perfectly achieved. At zero temperature the thermal equilibrium state degenerates into the ground state $|\phi_4\rangle$, which is a distorted singlet state. The unit fidelity indicates that in this protocol the reduction of the teleportation fidelity arises from this distortion can be completely modified.

The dependence of the average fidelity on the DM interaction $D$ at finite temperature is plotted in fig. (2). When the DM interaction is zero, one has $F_{sa} = F_{oa}$, which indicates that in this case the quality of the teleportation can not be improved to any extent. Combining the upper results for $T = 0$ and $D \neq 0$, it can be concluded that an additional unitary transformation can compensate the deviation of the ground state from the principal singlet state $|\psi^-\rangle$, **but** can not eliminate the influence caused by the mixture of the three excited states. Another notable phenomenon in fig. (2) is the different monotony of $F_{sa}$ and $F_{oa}$ on $D$, that is, while $F_{sa}$ decreases monotonously as $D$ increases, $F_{oa}$ increases instead. Although the DM interaction negatively impacts the teleportation quality in the standard protocol, it becomes an advantageous factor in the optimal protocol. This phenomenon is due to the influence of DM interaction on the ground state. On the one hand, the DM interaction introduces a relative phase in the ground state, which makes it deviate from the principal singlet state $|\psi^-\rangle$. Therefore, the teleportation quality will inevitably deteriorate **in the standard protocol**. On the other hand, the DM interaction decreases the energy level of the ground state. **As a result**, for a fixed temperature the ground state is more populated in the thermal equilibrium state, which is beneficial for the teleportation. In general cases, the first effect is more dominant. As a result, the teleportation fidelity is decreasing as a function of $D$ in the standard protocol,. In the optimal protocol, however, the first effect can be entirely eliminated, so that the second effect emerges.

In the above discussion we just focus on the Heisenberg *XXX* model in the presence of the DM anisotropy antisymmetric interaction. The effect of the DM interaction **can boil down to** an introduction of a relative phase in the ground state. Obviously, the different types of anisotropy interactions that have the similar **effects** can be modified in the same way. Besides, some anisotropy interactions or inhomogeneities of the system may affect the ground state in another way. Instead of introducing a relative phase, they may alter the amplitudes of $|10\rangle$ and $|01\rangle$ in the ground state. As proved below, however, this effect can not be compensated by any unitary transformations. According to the former discussions, the form of the optimal unitary transformation depends on the ground state only. So we can **simply** suppose the quantum channel is given by

$$|\chi\rangle = \frac{1}{\sqrt{2}}(\sqrt{1+u}|10\rangle - \sqrt{1-u}|01\rangle), \tag{27}$$

where $u$ is the deviation factor and satisfies $0 < |u| < 1$. Albeverio *et al.* [14, 19] showed that for the qubit system, when the state $\chi$ is used as the quantum channel, the optimal average teleportation fidelity can be written as

$$F_{oa} = \frac{2}{3}\mathbb{F}(\chi) + \frac{1}{3}, \tag{28}$$

where $\mathbb{F}$ is called the *fully entangled fraction* and defined as

$$\mathbb{F}(\chi) = \max\{\langle\psi^-|(I \otimes U^\dagger)\chi(I \otimes U)|\psi^-\rangle\}, \tag{29}$$

where $U$ is an arbitrary unitary transformation. Suppose $U$ takes the universal form in eq. (19), then the fully entangled fraction can be obtained as

$$\mathbb{F}(\chi) = \max\{\frac{1}{2}\cos^2\frac{\beta}{2}[1 + \sqrt{1-u^2}\cos(\alpha+\gamma)]\}. \tag{30}$$

In order to gain the maximum there should be $\cos^2\frac{\beta}{2} = \cos(\alpha+\gamma) = 1$, which yields $\alpha = \beta = \gamma = 0$ and $U = I$. In other words, **a superior unitary transformation that makes the fully entangled fraction as well as the average fidelity larger does not exist.** The major difference between $|\chi\rangle$ and $|\phi_4\rangle$ is that $|\phi_4\rangle$ is a maximal entangled state while $|\chi\rangle$ is not, though both of them have some deviation from the principal singlet state $|\psi^-\rangle$. When the ground state takes the form $|\chi_2\rangle = \frac{1}{\sqrt{2}}(\sqrt{1+u}|10\rangle - e^{i\vartheta}\sqrt{1-u}|01\rangle)$, which has a same entanglement degree as $|\chi\rangle$, the optimal unitary transformations for the Bell measurement outcomes $|\psi^-\rangle$, $|\psi^+\rangle$, $|\phi^-\rangle$, and $|\phi^+\rangle$ are $U_o(\vartheta)$, $U_o(\vartheta)\sigma^z$, $U_o(-\vartheta)\sigma^x$ and $U_o(-\vartheta)\sigma^y$, respectively. And the optimal teleportation fidelity is the same with the case that $|\chi\rangle$ is used as the quantum channel. From this we can draw the conclusion that for the optimal teleportation based on the Bell measurement the average fidelity is constrained by the entanglement degree. For a quantum channel that has a certain entanglement degree, there is an upper bound **for** the teleportation fidelity. The role of the optimal unitary transformation is to make the teleportation fidelity reach this upper bound while the entanglement degree of the quantum channel is fixed. And in the case that a unit fidelity is necessary while a maximal entangled quantum channel is not available, other protocols such as the so-called *probabilistic teleportation* [20-22], which achieves unit fidelity with a reduced probability instead of a lower fidelity with a unit probability, should be applied.

In conclusion, we have investigated the optimal teleportation based on the Bell measurement via a two-qubit Heisenberg *XXX* chain in the presence of DM anisotropic antisymmetric interaction. We give the explicit expression of the optimal unitary transformations and show that they depend on the ground state only. It is found that in this protocol the fidelity reduction caused by the DM interaction can be completely eliminated at zero temperature. At finite temperatures the DM interaction can enhance the teleportation fidelity, in contrast to its effect in the standard protocol. We also discuss the effect of other anisotropies and show that the deviation of the amplitude in the ground state can not be compensated in this protocol.

**ACKNOWLEDGEMENT**

This work is supported by the National Science Foundation of China under Grant No. 10604053 & 10874013 and the Beihang Lantian Project. Sh-Sh Li also acknowledges the support of National Basic Research Program of China (973 Program) Grant No. G2009CB929300 and the National Natural Science Foundation of China under Grant Nos. 60821061 & 60776061. Ahmad Abliz also acknowledges the support of National Science Foundation of China under Grant No. 10664004.

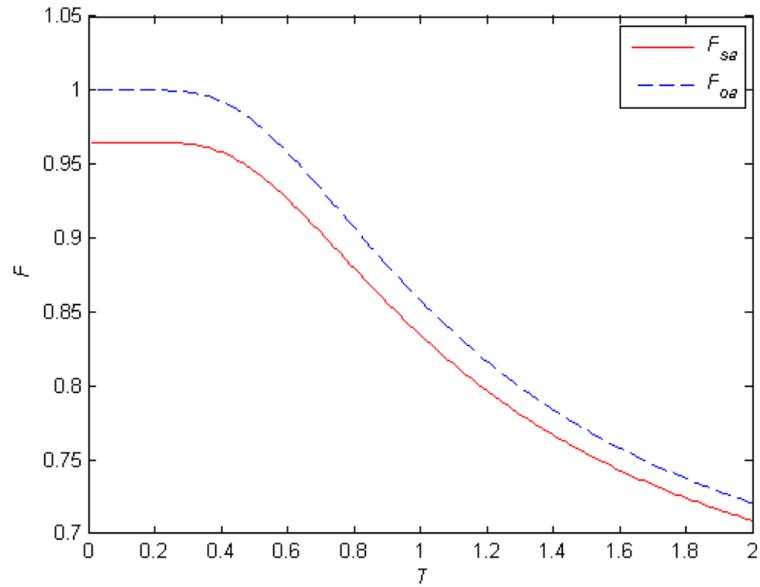

Fig. 1. (Color online) $T$-dependence of the average teleportation fidelity with $J$=1 and $D$=0.5.

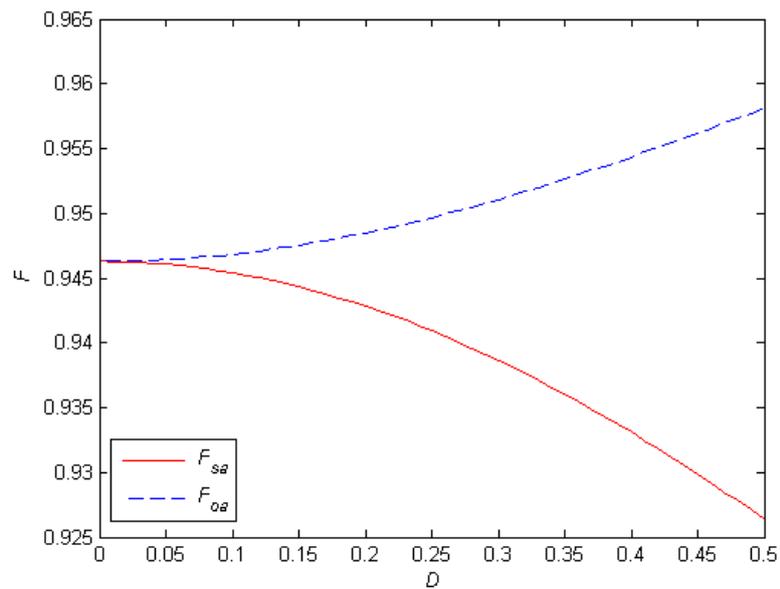

Fig. 2. (Color online) $D$-dependence of the average teleportation fidelity with $J$=1 and $T$=0.6.